\documentclass[pra,twocolumn,a4paper,superscriptaddress]{revtex4-1}
\usepackage{amssymb}
\usepackage{amsmath}
\usepackage{graphicx}
\usepackage{bm}
\usepackage{times}
\usepackage{color}
\definecolor{blue}{RGB}{25,51,200}
\definecolor{green}{RGB}{51, 150, 25}
\definecolor{red}{RGB}{200,25,25}
\usepackage[colorlinks,citecolor=green,linkcolor=blue,urlcolor=red]{hyperref}
\usepackage{dsfont}

\renewcommand{\d}{\mathrm{d}}
\def\mD{\mathcal{D}}
\def\mH{\mathcal{H}}

\begin{document}

\title{Dissipative vs. Conditional Generation of Gaussian Entanglement and Spin Squeezing}

\author{Denis V. Vasilyev}
\email{denis.vasilyev@itp.uni-hannover.de}
\affiliation{Institute for Theoretical Physics, Institute for Gravitational Physics (Albert Einstein Institute), Leibniz University Hannover, Callinstra\ss{}e 38, 30167 Hannover, Germany}
\author{Christine A.\ Muschik}
\affiliation{ICFO-Institut de Ci\`{e}ncies Fot\`{o}niques, Mediterranean Technology Park, 08860 Castelldefels
(Barcelona), Spain.}
\author{Klemens Hammerer}
\affiliation{Institute for Theoretical Physics, Institute for Gravitational Physics (Albert Einstein Institute), Leibniz University Hannover, Callinstra\ss{}e 38, 30167 Hannover, Germany}

\date{\today}

\begin{abstract}
Spin squeezing of collective atomic spins can be achieved conditionally via probing with light and subsequent homodyne detection, as is done in a Quantum Nondemolition measurement. Recently it has been shown that squeezing can also be created unconditionally by a properly designed dissipative dynamics. We compare the two approaches in a Gaussian description, and optimize over all Gaussian light-matter interactions. We find that in the optimal unconditional scheme based on dissipation the level of squeezing scales with optical depth as $d^{-1/2}$. In contrast, the optimal conditional scheme based on measurement of light -- which in fact is \emph{not} a Quantum Nondemolition measurement --  can provide squeezing which scales as $d^{-1}$. Our results apply directly also to the creation of entanglement in the form of non-local spin squeezing of two atomic ensembles.
\end{abstract}

\maketitle

\section{Introduction}\label{sec:intro}


Engineered dissipative dynamics can stably generate nonclassical states and nontrivial dynamics of quantum systems, as was shown in recent theoretical \cite{DallaTorre2013,Clark2003,Kastoryano2011,Stannigel2011,Parkins2006,Zheng2010,Muschik2011,Muschik2012,Watanabe2012,Yang2012a,Tan2013,Diehl2008,Kraus2008,Verstraete2009,vollbrecht_entanglement_2011,Muller2011,Genoni2012} and experimental \cite{Krauter2010,Barreiro2011} work. Dissipative generation of spin squeezing \cite{DallaTorre2013}, entangled states of single atoms \cite{Clark2003,Kastoryano2011,Stannigel2011}, atomic ensembles \cite{Parkins2006,Zheng2010,Krauter2010,Muschik2011,Muschik2012,Watanabe2012,Yang2012a} or micromechanical oscillators \cite{Tan2013,Tan2013a}, and quantum states of many body systems \cite{Diehl2008,Kraus2008} is possible, and even quantum computation \cite{Verstraete2009}, quantum communication \cite{vollbrecht_entanglement_2011}, and quantum simulations \cite{Muller2011,Barreiro2011} can be performed purely based on dissipation.  Another method for engineering of quantum states and dynamics, which likewise rests upon the coupling between the system to be controlled and its environment, is continuous measurement and feedback control \cite{parthasarathy_introduction_1992,carmichael_open_1993,wiseman_quantum_2009,barchielli_quantum_2009,Thomsen2002,Thomsen2002_2}. Here the state or dynamics of the system is conditioned upon a measurement performed on the environment, and may be controlled via feedback depending on the measurement outcomes. Continuous measurement and feedback control is a well established  method in an equally broad range of physical systems \cite{bushev_feedback_2006,Kubanek2009,Koch2010,Sayrin2011,haroche_exploring_2006,smith_efficient_2006,Chaudhury2007,Krauter2010,arcizet_high-sensitivity_2006,Poggio2007,Montinaro2012,corbitt_optical_2007,mow-lowry_cooling_2008,Abbott2009}. This naturally raises the question of which methods work best for a given task or figure of merit.

In this article we address this question for the case of spin squeezing \cite{Ma2011} and spin squeezing based entanglement of atomic ensembles \cite{hammerer_quantum_2010}. The creation of spin squeezing can be considered as the prime example for measurement and feedback based quantum control. It relies on the optical probe of a collective atomic spin by off-resonant light, homodyne detection of light, and an appropriate feedback on the atomic spin \cite{Kuzmich2000}. The light-matter interaction is thereby commonly tailored such that a quantum nondemolition measurement (QND) \cite{Braginsky1980,Grangier1998} of the atomic spin is realized. Squeezing of atomic spins via (continuous or pulsed) QND probes with and without feedback was demonstrated in \cite{Appel2009,Chen2011,Inoue,Kuzmich2000,Schleier-Smith2010,Sewell2012,Sewell2013,Takano2009}. When squeezing of two collective spins is realized in this way, entanglement in the form of Einstein-Podolsky-Rosen (EPR) squeezing is established among the two atomic ensembles \cite{hammerer_quantum_2010}, as was done in \cite{Julsgaard2001}. In the alternative context of dissipative quantum state engineering it was shown recently that a careful choice of the light-matter interaction can provide a spin squeezed or entangled state in steady state without applying any measurement on the light and feedback on the atoms \cite{Muschik2011,Muschik2012}. This was realized experimentally soon after \cite{Krauter2010}, albeit with the help of moderate feedback compensating for experimental imperfections.

Prompted by these theoretical studies and experiments we investigate here how the light-matter interaction has to be tailored in order to achieve a maximal amount of spin squeezing or entanglement. We consider and compare all three relevant scenarios: unconditional, dissipative state generation, measurement based conditional state generation, and unconditional measurement and feedback assisted state generation. In each case we optimize over the class of Gaussian light-matter interactions (\textit{i.e.} those which are quadratic in (collective) creation and annihilation operators, see below). Crucially, we take into account decoherence induced by spontaneous emission due to the off-resonant light probe. We do this on the basis of a minimal, yet microscopically justified Gaussian model \cite{Vasilyev2012}. This approach allows us to relate the spin noise reduction or entanglement achievable under optimized conditions eventually to the optical depth $d$ of the atomic ensemble~\footnote{The optical depth $d$ is the number of atoms in a column along the propagation direction of light with diameter of a wavelength.}.

\begin{figure}[t]
  \includegraphics[width=0.9\columnwidth]{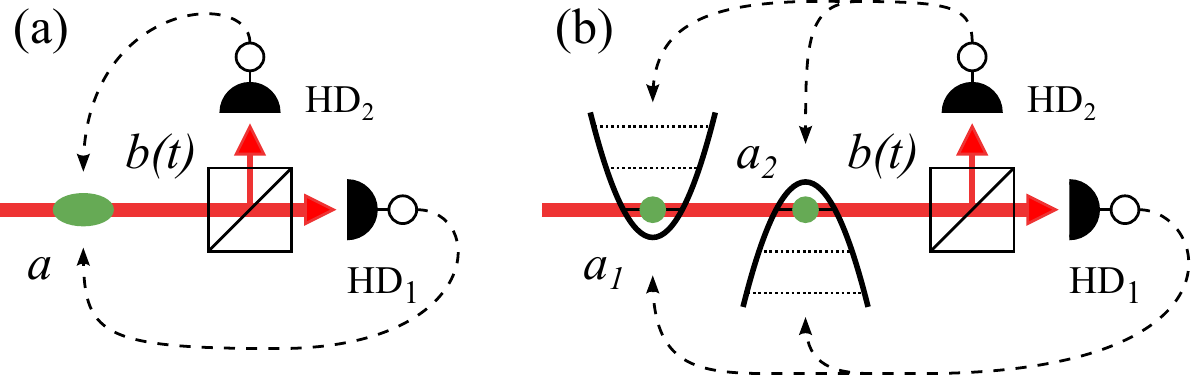}
  \caption{(Color online) One (a) or two (b) bosonic modes (collective atomic spins) 
  couple sequentially to 1D field, which might be subject to homodyne detection after the interactions.}\label{fig:setup}
\end{figure}

Our results are as follows: First, for unconditional, dissipative state generation the optimal (EPR) squeezing scales as $d^{-1/2}$. This is the same scaling as is achievable with a QND measurement \cite{Hammerer2004}. While the scaling with optical depth is the same in both cases a QND measurement provides squeezing which is twice as large for the same parameters. However, QND squeezing is prepared conditionally and becomes unconditional only when feedback is applied. Thus, dissipative generation of squeezing spares the need for feedback at the cost of a 50\% reduced level of squeezing. Second, measurement induced squeezing can exhibit a rather remarkable scaling of $d^{-1}$ for the regime of moderate optical depths. In this most relevant regime we thus predict an enhancement of $d^{-1/2}$ in comparison  to the scaling for QND squeezing. For large optical depths the optimal scaling eventually approaches $d^{-1/2}$ again. The enhancement comes at the cost of the necessity of applying feedback in order to stabilize the spin state as the conditional state is unstable (in contrast to the case of QND squeezing). In each case we determine the optimal light matter interaction as a sum of beam splitter and two mode squeezing dynamics, with dominating beam splitter contribution for the dissipative generation, and dominating two mode squeezing for the measurement based scheme.

Our results are formally derived by means of Gaussian quantum stochastic master equations and corresponding 
equations for displacement vectors and covariance matrices (\textit{i.e.} the first and second moments) \cite{Edwards2005, Wiseman2010}. We supply the basics and some useful relations for this formalism in Appendix~\ref{app:belavkin}.

The paper is organised as follows: In Sec.~\ref{sec:model} we introduce a model describing atomic ensembles interacting with an off-resonant probe light. In the next Section~\ref{sec:singlespin} we derive the stochastic master equation for a single bosonic mode probed by light. After that the corresponding equations for variances of the oscillator quadratures are introduced and solved for conditional and unconditional dynamics. In Sec.~\ref{sec:twospins} we consider two bosonic modes (spin ensembles) coupled to a 1D field in a cascaded fashion. We show that the evolution of EPR modes of the two oscillators coincides with the dynamics of the single bosonic mode considered in previous section. The discussion of our main results for spin squeezing and entanglement is given in Section~\ref{sec:discussion}.

\section{Model}\label{sec:model}

We want to describe a general interaction of an atomic ensembles of spin polarised atoms with a 1D light field which leads to spin squeezing or generation of entanglement. The widely used procedure for a macroscopically big collective spin is to approximate the spin component along its axis of polarisation with its mean value $J_{x}\approx\langle J_{x}\rangle$. Rescaling of the rest of the spin components provides the canonical position $X=J_{y}/\sqrt{\langle J_{x}\rangle}$ and momentum $P=J_{z}/\sqrt{\langle J_{x}\rangle}$ operators with a proper computational relation $[X,P]=i$. The rigorous description of this procedure is given by the Holstein-Primakoff transformation \cite{Holstein:1940fk}. It allows us to work with bosonic modes and obtain results which can be applied to macroscopic ensembles of spin polarised atoms.

The optical probe will inevitably cause decoherence in the spin systems, which can be taken into account by introducing decay rates and effective occupation numbers of the environment:
	\begin{align*}
	\dot X &= -\gamma_{x}X + f_{x}(t), & \langle f_{x}(t)f_{x}(t')\rangle &= n_{x}\delta(t-t'),\\
	\dot P &= -\gamma_{p}P + f_{p}(t), &     \langle f_{p}(t)f_{p}(t')\rangle &=n_{p}\delta(t-t').
	\end{align*}
The relevant decay rates and noise terms due to the interaction of atoms with the reservoir of electromagnetic modes can be calculated form first principles, as was done in \cite{Vasilyev2012}. Decoherence from other sources (if present) can be added independently. In order to keep the following discussion conceptually simple we will assume equal decay rates $\gamma_{x}=\gamma_{y}=\gamma$ and occupation numbers $n_{x}=n_{y}=n$. Our approach can be easily adapted in order to account for more general situations. The equivalent description of the decoherence process in terms of the master equation is given by
	\begin{equation}
	\dot\rho = \gamma(n+1)\mD[a]\rho + \gamma n\mD[a^{\dag}]\rho,
	\end{equation}
where the Lindblad term is defined as $\mD[x]\rho=x\rho x^\dagger -\tfrac12\{x^\dagger x,\rho\}_+$.

We consider a setup as shown in Fig.~\ref{fig:setup}. A 1D field is coupled to one or two localized bosonic modes positioned at points $z_1$ and $z_2$. In the latter case the field thus interacts sequentially with the two localized modes at times $t_1=z_1/c$ and $t_2=z_2/c>t_1$. After the interaction the field can be homodyne detected. We are interested in the stationary state of the localized bosonic modes, and in particular in its squeezing and entanglement properties. Stationarity is thereby understood with respect either to the purely dissipative, unconditional dynamics or to the dynamics conditioned on the homodyne detection.

\section{A single oscillator coupled to a 1D field}\label{sec:singlespin}

\subsection*{Conditional Master Equation}

 We consider first a system with one localised bosonic mode coupled to a 1D field as shown in Fig.~\ref{fig:setup}a. The 1D field is described by bosonic operators in frequency space $b_\omega$, or in time
 \[
 b(t)=\int\frac{\mathrm{d}\omega}{\sqrt{2\pi}}\,b_\omega e^{i\omega t},
 \]
 where $t=z/c$. The corresponding quadrature operators are $x(t)=[b(t)+b(t)^\dagger]/\sqrt{2}$ and $p(t)=-i[b(t)-b(t)^\dagger]/\sqrt{2}$, and carry dimension Hz$^{1/2}$. The oscillator is described by a dimensionless bosonic operator $a$ with corresponding canonical operators $X=(a+a^\dagger)/\sqrt{2}$ and $P=-i(a-a^\dagger)/\sqrt{2}$.

 We assume a general quadratic interaction of the field with the oscillator. Without loss of generality \cite{Kraus2003} any such interaction can be written as
 \[
   H^\mathrm{int}=\sqrt{g}\big[\alpha Xx(t) + \beta Pp(t)\big],
 \]
 where $g$ has dimension Hz and determines the strength of the coupling, and the coefficients $\alpha$ and $\beta$ can be parameterized as
 \begin{align}\label{eq:alphabeta}
   \alpha&=\cos(\theta) & \beta&=\sin(\theta)
 \end{align}
 The ratio of these coefficients determines the qualitative character of the coupling. When written in terms of creation and annihilation operators the Hamiltonian becomes
 \[
   H^\mathrm{int}=\sqrt{g} \left[s^\dagger b(t) +  s b^\dagger(t)\right].
 \]
 where
 \begin{equation*}
   s = \frac1{\sqrt2}(\alpha X + i\beta P)
   = \frac{\alpha+\beta}{2}a + \frac{\alpha-\beta}{2}a^\dagger.
 \end{equation*}
Thus, for $\theta=0$ the Hamiltonian describes a QND interaction $\sim Xx(t)$, while for $\theta=\pi/4$ it is a beam splitter interaction $\sim a b^\dagger(t)+a^\dagger b(t)$, and for $\theta=-\pi/4$ it is a two mode squeezing interaction $\sim a b(t)+a^\dagger b^\dagger(t)$. Other values of $\theta$ cover any quadratic interaction in between.

 The full Hamiltonian for the oscillator and the 1D field is
 \begin{align*}
      H&=H + H_\mathrm{bath} + H^\mathrm{int}.
 \end{align*}
 where $H$ is a local Hamiltonian for the localized mode, and $H_\mathrm{bath}=\int\mathrm{d}\omega\,\omega b^\dagger_\omega b_\omega$. Eliminating the bath in a Born-Markov approximation yields a master equation 
 \begin{align}\label{eq:single_meq}
        \dot\rho& = -i[H,\rho]+g \mD[s]\rho
 \end{align}
 where the Lindblad term is defined as $\mD[x]\rho=x\rho x^\dagger -\tfrac12\{x^\dagger x,\rho\}_+$.

 To the master equation \eqref{eq:single_meq} we can now add other processes relevant to the dynamics, as explained in Sec.~\ref{sec:model}: First, we assume that the oscillator couples at a rate $\gamma$ to a thermal reservoir, causing in thermal equilibrium a mean occupation number of $n$, and therefore implies a thermal decoherence rate $\gamma^{\mathrm{th}}=\gamma{n}$. Second, if the transmitted field is subject to homodyne detection of both conjugate light quadratures (after being split on a beamsplitter with reflectivity $\epsilon$, cf. Fig.~\ref{fig:setup}) the conditional state evolves according to the conditional stochastic master equation
 \begin{align}\label{eq:singleSMEQ}
        \mathrm{d}\rho = &-i[H,\rho]\mathrm{d}t + g \mD[s]\rho\mathrm{d}t\notag\\
        &+\sqrt{g(1-\epsilon)}\mH[se^{i\phi}]\rho \mathrm{d}W_{1} + \sqrt{g\epsilon}\mH[ise^{i\phi}]\rho \mathrm{d}W_{2}\notag\\
        &+\left\{\gamma({n}+1) \mD[a]\rho
        +\gamma{n}\mD[a^\dagger]\rho\right\}\mathrm{d}t,
 \end{align}
 where $\mH[x]\rho=(x-\langle x\rangle)\rho+\rho(x^\dagger-\langle x^\dagger\rangle)$ and $\mathrm{d}W_{i}$ is a Wiener increment of zero mean and $\mathrm{d}W_{i}^2=\mathrm{d}t$. $\phi$ is the phase of the local oscillator in the homodyne detection. By convention $\phi=0$ corresponds to a measurement of the phase quadrature $p(t)$.

\subsection*{Steady State Variances}

   For Gaussian states the stochastic master equation \eqref{eq:singleSMEQ} implies a time evolution which is described by a stochastic linear differential equation for the conditional mean values $\langle X(t)\rangle$ and $\langle P(t)\rangle$, and a \textit{deterministic} nonlinear differential equation for the conditional variances
   \begin{align}\label{eq:vars}
     V_\mathrm{c}&=\Delta X^2, & U_\mathrm{c}&=\Delta P^2,
   \end{align}
   and covariances $C_\mathrm{c}=\langle XP+PX\rangle-2\langle X\rangle\langle P\rangle$. 
   In \cite{Edwards2005,Genoni2012} compact formulas were derived in order to pass from a general stochastic master equation such as \eqref{eq:singleSMEQ} to the corresponding equations for $V_\mathrm{c}$, $U_\mathrm{c}$, and $C_\mathrm{c}$. In Appendix~\ref{app:belavkin} we provide a summary of this extremely useful formalism.

   For an optimal choice $\phi=0$ of the local oscillator phase (giving rise to the strongest squeezing) and a beamsplitter reflectivity $\epsilon$ the \emph{conditionally} squeezed variance $V_\mathrm{c}$ and the \emph{conditionally} unsqueezed variance $U_\mathrm{c}$ obey the nonlinear equations of motion
   \begin{align}\label{eq:EoMVc}
      \dot{V}_\mathrm{c} = &-[\gamma - g(1-2\epsilon)\alpha\beta]V_\mathrm{c} - g(1-\epsilon)\alpha^{2}V_\mathrm{c}^2 \notag\\
      &+\gamma(2n+1) + g\epsilon \beta^{2},\\
      \dot{U}_\mathrm{c} = &-[\gamma + g(1 - 2\epsilon)\alpha\beta]U_\mathrm{c} - g\epsilon\beta^{2}U_\mathrm{c}^2 \notag\\
      &+\gamma(2n+1) + g(1-\epsilon)\alpha^{2},
   \end{align}
   as follows from Eqs.~\eqref{eq:Belavkin} in Appendix~\ref{app:belavkin}.
   We suppress the corresponding equation of motion for the conditional covariance $C_\mathrm{c}$, as it will vanish in steady state for the present case. The conditional variances have the steady state solutions
   \begin{align}\label{eq:Vc}
      V_\mathrm{c} &= \frac1{2(1-\epsilon)g\alpha^{2}}\big\{(1-2\epsilon)g\alpha\beta - \gamma\notag\\
      & + \sqrt{(\gamma-g\alpha\beta )^{2} + 4\gamma g\alpha[(2n+1)(1-\epsilon)\alpha + \epsilon\beta]} \big\},\\
      \label{eq:Uc}
      U_\mathrm{c}&=\frac{1}{2\epsilon g\beta^{2}}\big\{-(1-2\epsilon)g\alpha\beta - \gamma \notag\\
      & + \sqrt{(\gamma + g\alpha\beta)^{2} + 4\gamma g\beta\epsilon[(2n+1)\beta-\alpha]}\big\}.
   \end{align}

   The result for the conditional variances in steady state \eqref{eq:Vc} and \eqref{eq:Uc} can be compared to the unconditional variances $V_\mathrm{u}$ and $U_\mathrm{u}$, which can be obtained by solving \eqref{eq:singleSMEQ} neglecting the stochastic term proportional to the Wiener increments. Applying again formulas \eqref{eq:Belavkin} the master equation implies for the unconditional squeezed and unsqueezed variances
   \begin{align}\label{eq:EoMVu}
      \dot{V}_\mathrm{u} &= -(\gamma + g\alpha\beta)V_\mathrm{u} + g\beta^{2} + \gamma(2n+1),\\
      \dot{U}_\mathrm{u} &= -(\gamma + g\alpha\beta)V_\mathrm{u} + g\alpha^{2} + \gamma(2n+1)
   \end{align}
   with the steady state solutions
   \begin{align}\label{eq:Vu}
      V_\mathrm{u} &= \frac{g\beta^{2} + \gamma(2n+1)}{\gamma + g\alpha\beta},\\
      \label{eq:Uu}
      U_\mathrm{u} &= \frac{g\alpha^{2} + \gamma(2n+1)}{\gamma + g\alpha\beta}.
   \end{align}
   Equations \eqref{eq:Vc}, \eqref{eq:Uc}, and \eqref{eq:Vu} are the main results of this section.

   Instead of asking for the steady state corresponding to a continuous wave drive and a continuous detection of light, we can also ask for the solution to the equation of motion for the conditional variance $V_\mathrm{c}$ for \emph{finite} times corresponding to driving the system with a pulse. Integrating \eqref{eq:EoMVc} for a time $\tau$, measuring only one quadrature of light ($\epsilon=0$), neglecting any losses ($\gamma=0$), and selecting a QND interaction ($\theta=0$) with initial condition $V_\mathrm{c}(0)=1$ yields $V_\mathrm{c}(\tau)=1/(1+g\tau)$. We can compare this to the known result for QND squeezing, which is given by $V^\mathrm{QND}_\mathrm{c}(\tau)=1/(1+\kappa^2)$ where $\kappa^2=d\eta$ with optical depth $d$ and $\eta=\gamma\tau$ the degree of atomic depumping from a pulse with duration $\tau$, see \emph{e.g.}~\cite{hammerer_quantum_2010}. We can therefore conclude that
   \begin{align}\label{eq:g}
   g=d\gamma.
   \end{align}
   Since the expressions for conditional \eqref{eq:Vc}, \eqref{eq:Uc} and unconditional \eqref{eq:Vu}, \eqref{eq:Uu} variances in steady state depend only on the ratio $\gamma/g$ the results of the substitution $g= d\gamma$ become independent of the decay rate, and depend only on the optical depth, the effective mean thermal occupation, and the parameters $\epsilon$ and $\theta$ characterzing the measurement and the light matter interaction. Substituting \eqref{eq:g} and \eqref{eq:alphabeta} into \eqref{eq:Vc}, \eqref{eq:Uc} and \eqref{eq:Vu}, \eqref{eq:Uu} we arrive at the final results given in Sec.~\ref{sec:discussion}, Eqs.~\eqref{eq:VuODtheta} and \eqref{eq:VcODtheta}. Before we enter into the discussion of these results we will show that they apply directly also to the EPR squeezing of two bosonic modes.


    \section{Cascaded Counterrotating Configuration}\label{sec:twospins}
The preparation of a squeezed state for an oscillator is closely related to the production of an entangled state of two oscillators. Essentially one needs to prepare EPR modes of the oscillators in a squeezed state. In order to achieve this we consider two oscillators coupled to the same field mode in a cascaded fashion as shown in Fig.~\ref{fig:setup}b. The full Hamiltonian for the cascaded system of the two localized modes and the 1D field is
 \begin{align*}
      H&=H_1+H_2+H_\mathrm{bath}+H_1^\mathrm{int}+H_2^\mathrm{int}.
 \end{align*}
 where $H_i$ are local Hamiltonians for the two modes, and
 \[
   H_i^\mathrm{int}=\sqrt{g_i}\big[\alpha_i X_ix(t_i) + \beta_i P_ip(t_i)\big],
 \]
 with $t_2>t_1$. To ease notation we assume couplings of equal strength $g_{i=1,2}=g$. Eliminating the bath in a Born-Markov approximation yields a cascaded systems master equation \cite{Kolobov1987, Gardiner1993, Carmichael1993, Stannigel2010, Stannigel2011}
 \begin{align}\label{eq:meq}
        \dot\rho&=-i\left[H_1+H_2-i\frac{g}{2}(s_2^\dagger s_1-s^\dagger_1 s_2),\rho\right]+g \mD[s_1+s_2]\rho
 \end{align}
 where the Lindblad term depends on a \emph{collective} jump term $s_1+s_2$. The effective interaction among the two modes mediated via the 1D field is
 \begin{align*}
     -i\frac{g}{2}(s_2^\dagger s_1-s^\dagger_1 s_2) 
      =&-i\frac{g}{2}\left[\frac{\alpha_{2}\beta_1-\alpha_{1}\beta_2}{2}\left(a_{1}a_{2}-a_{1}^\dagger a_{2}^\dagger\right)\right.\\
      &\left.+\frac{\alpha_{2}\beta_1+\alpha_{1}\beta_2}{2}\left(a_{1}a_{2}^\dagger-a_{1}^\dagger a_{2}\right)\right].
 \end{align*}
 Note that if both modes couple in the same way to the 1D field, $\alpha_{1}=\alpha_{2}$, $\beta_1=\beta_2$, this mediated interaction is a pure beam splitter like coupling.

 As we did above with a single oscillator, we can now add to the cascaded systems master equation \eqref{eq:meq} other processes relevant to the dynamics: First, we assume both modes couple at a rate $\gamma_i$ to a thermal reservoir causing in thermal equilibrium a mean occupation number of ${n}_i$, and therefore implying a thermal decoherence rate $\gamma_i^{\mathrm{th}}=\gamma_i{n}_i$. Second, if the transmitted field is subject to homodyne detection the conditional state evolves according to the conditional stochastic master equation
 \begin{align*}
        \mathrm{d}\rho = &-i\left[H_1+H_2-i\frac{g}{2}(s_2^\dagger s_1-s^\dagger_1 s_2),\rho\right]\mathrm{d}t\\
        &+g \mD[s_1+s_2]\rho\mathrm{d}t+\sqrt{g(1-\epsilon)}\mH[(s_1+s_2)e^{i\phi}]\rho \mathrm{d}W_{1}\\
        &+\sqrt{g\epsilon}\mH[i(s_1+s_2)e^{i\phi}]\rho \mathrm{d}W_{2}\\
        &+\sum_i\left\{\gamma({n}_i+1) \mD[a_i]\rho
        +\gamma{n}_i\mD[a^\dagger_i]\rho\right\}\mathrm{d}t.
 \end{align*}

 We now specialize this fairly general model. First, we assume the ``Copenhagen setup'' of counterrotating harmonic oscillators in which a broad range of quantum information protocols were performed \cite{hammerer_quantum_2010}. In this setup the local Hamiltonians are
 \begin{align*}
        H_1+H_2&=\omega a_1^\dagger a_1-\omega a_2^\dagger a_2.
 \end{align*}
 such that one of the ensembles effectively realizes an oscillator of negative mass. In a rotating frame with respect to $H_1+H_2$ one finds
 \[
      s_1(t)+s_2(t)=s_+e^{-i\omega t}+s_-e^{i\omega t}
 \]
 where
 \begin{align*}
      s_+&=\frac{\alpha_{1}+\beta_1}{2}a_1+\frac{\alpha_{2}-\beta_2}{2}a_2^\dagger \\
      s_-&=\frac{\alpha_{2}+\beta_2}{2}a_2+\frac{\alpha_{1}-\beta_1}{2}a_1^\dagger.
 \end{align*}
 Assuming $g,\,\gamma_i^\mathrm{th}\ll\omega$ we can perform a rotating wave approximation such that the conditional master equation becomes
 \begin{align*}
        \mathrm{d}\rho = &-i\left[\frac{-ig(\alpha_{2}\beta_1-\alpha_{1}\beta_2)}{4}(a_{1}a_{2}-a_{1}^\dagger a_{2}^\dagger),\rho\right]\mathrm{d}t\\
        &+\frac{g}{2}\mD[s_++s_-]\rho\mathrm{d}t+\frac{g}{2}\mD[s_+-s_-]\rho\mathrm{d}t\\
        &+\sqrt{\frac{g(1-\epsilon)}{2}}\mH[(s_++s_-)e^{i\phi}]\rho \mathrm{d}W_{c1}\\
        &+\sqrt{\frac{g\epsilon}{2}}\mH[i(s_++s_-)e^{i\phi}]\rho \mathrm{d}W_{c2}\\
        &+\sqrt{\frac{g(1-\epsilon)}{2}}\mH[-i(s_+-s_-)e^{i\phi}]\rho \mathrm{d}W_{s1}\\
        &+\sqrt{\frac{g\epsilon}{2}}\mH[(s_+-s_-)e^{i\phi}]\rho \mathrm{d}W_{s2}\\
        &+\sum_i\left\{\gamma_i({n}_i+1) \mD[a_i]\rho
        +\gamma_i{n}_i\mD[a^\dagger_i]\rho\right\}\mathrm{d}t.
 \end{align*}
 Here $\mathrm{d}W_{c1(2)}$ and $\mathrm{d}W_{s1(2)}$ are independent Wiener increments, corresponding to noise in sine and cosine modulation modes at the sideband frequencies $\pm\omega$ for the $x$ ($p$) quadrature. For details we refer to \cite{Wiseman2010} (Sec.~4.5).

 For an entirely symmetric situation $\alpha_{i}=\alpha$, $\beta_i=\beta$, $\gamma_i=\gamma$, and ${n}_i={n}$ for both $i=1,2$, we get
 \begin{align*}
        s_++s_- &= \alpha X_+ + i\beta P_+\\
        s_+-s_-&=i(\alpha P_- - i\beta X_-)
 \end{align*}
 where the EPR operators are $X_\pm=(X_1\pm X_2)/\sqrt{2}$ and $P_\pm=(P_1\pm P_2)/\sqrt{2}$. In this case the dynamics factorizes in these modes, and the reduced density operators $\rho_\pm$ for the EPR modes fulfill the stochastic master equations
      \begin{align}\label{eq:cMEQ+}
        \mathrm{d}\rho_+=\frac{g}{2}&\mD[\alpha X_++i\beta P_+]\rho_+\mathrm{d}t\\
        &+\sqrt{\frac{g(1-\epsilon)}{2}}\mH[(\alpha X_++i\beta P_+)e^{i\phi}]\rho_{+} \mathrm{d}W_{c1}\nonumber\\
        &+\sqrt{\frac{g\epsilon}{2}}\mH[i(\alpha X_++i\beta P_+)e^{i\phi}]\rho_{+} \mathrm{d}W_{c2}\nonumber\\
        &+\gamma({n}+1) \mD[a_+]\rho_{+}\mathrm{d}t
        +\gamma{n}\mD[a^\dagger_+]\rho_{+}\mathrm{d}t,\nonumber\\
        \label{eq:cMEQ-}
        \mathrm{d}\rho_-=\frac{g}{2}&\mD[\alpha P_--i\beta X_-]\rho_-\mathrm{d}t\\
        &+\sqrt{\frac{g(1-\epsilon)}{2}}\mH[-i(\alpha P_--i\beta X_-)e^{i\phi}]\rho_- \mathrm{d}W_{s1}\nonumber\\
        &+\sqrt{\frac{g\epsilon}{2}}\mH[(\alpha P_--i\beta X_-)e^{i\phi}]\rho_- \mathrm{d}W_{s2}\nonumber\\
        &+\gamma({n}+1) \mD[a_-]\rho_{-}\mathrm{d}t\nonumber
        +\gamma{n}\mD[a^\dagger_-]\rho_{-}\mathrm{d}t,\nonumber
      \end{align}
      where $a_\pm=(X_\pm+iP_\pm)/\sqrt{2}$. One can see that the equations \eqref{eq:cMEQ+} and \eqref{eq:cMEQ-} obtained for the two EPR modes ($+$ and $-$) are equivalent to the stochastic master equation for a single oscillator considered in Sec.~\eqref{sec:singlespin}, cf.~\eqref{eq:singleSMEQ}. Therefore, the steady state solutions for the single oscillator's variances \eqref{eq:Vc}, \eqref{eq:Uc}, \eqref{eq:Vu} describe the squeezing of EPR modes as well. Entanglement of the two oscillators is established if the sum of the two squeezed EPR variances falls below two units of spin projection noise (that is, $2\times1/2$ in our convention).

\section{Discussion}\label{sec:discussion}

Now we come to the discussion of the results obtained for the squeezing of a collective spin or entanglement of two of them. As explained at the end of Sec.~\ref{sec:singlespin} the expressions for the unconditional \eqref{eq:Vu} and conditional \eqref{eq:Vc} variances in terms of the interaction parameter $\theta$ and the optical depth $d$ read
	\begin{align}\label{eq:VuODtheta}
      V_\mathrm{u} = \frac{2n+1+d\sin^{2}\!\theta}{1 + d\sin(2\theta)/2},
	\end{align}
	\begin{align}\label{eq:VcODtheta}
      V_\mathrm{c}&=\frac1{2(1-\epsilon)}\left\{(1+\epsilon)\tan\theta - \frac{1}{d\cos^{2}\!\theta} \right.\notag\\
      &\left.+\sqrt{\left(\frac{1}{d\cos^{2}\!\theta} -\tan\theta\! \right)^{2} \!\!\! + \frac{4[(2n+1)(1-\epsilon)+\epsilon\tan\theta]}{d\cos^{2}\!\theta}} \right\}.
   \end{align}

   \paragraph*{Unconditional variance}
   \begin{itemize}
   \item
   The unconditional steady state exists only provided $\theta>\theta_{c}$ where the critical interaction parameter is given by $\gamma + g\sin(2\theta_{c})/2=0$, or \begin{align}\label{eq:thetac}
   \theta_{c}=-\frac12\arcsin(2/d).
   \end{align}
   Otherwise the system becomes dynamically unstable, see \eqref{eq:EoMVu}. However, this is not the case for the conditional state which can be stabilised for any interaction parameter $\theta$ by the measurement of both quadratures of the output light ($\epsilon\neq0$ or 1) and a feedback (see below).

	\item
   The unconditional variance \eqref{eq:VuODtheta} exhibits a tradeoff with respect to $\theta$, as is shown in Fig.~\ref{fig:Vu}. 
   The minimum of the variance (maximum of entanglement) is achieved for the optimal interaction parameter $\theta^\mathrm{opt}_\mathrm{u}$ given in the Appendix~\ref{appOptimalTheta}, and in general requires an interaction dominated by the beam splitter dynamics. In the limit of high optical depth it reads
   \begin{equation}\label{eq:OptimalThetau}
   \theta^\mathrm{opt}_\mathrm{u} \simeq \sqrt{\frac{2n+1}d}.
   \end{equation}
   The exact value of $\theta^\mathrm{opt}_\mathrm{u}$ is shown in Fig.~\ref{fig:optTheta} as a function of optical depth.
\begin{figure}[t] 
   \centering
   \includegraphics[width=0.9\columnwidth]{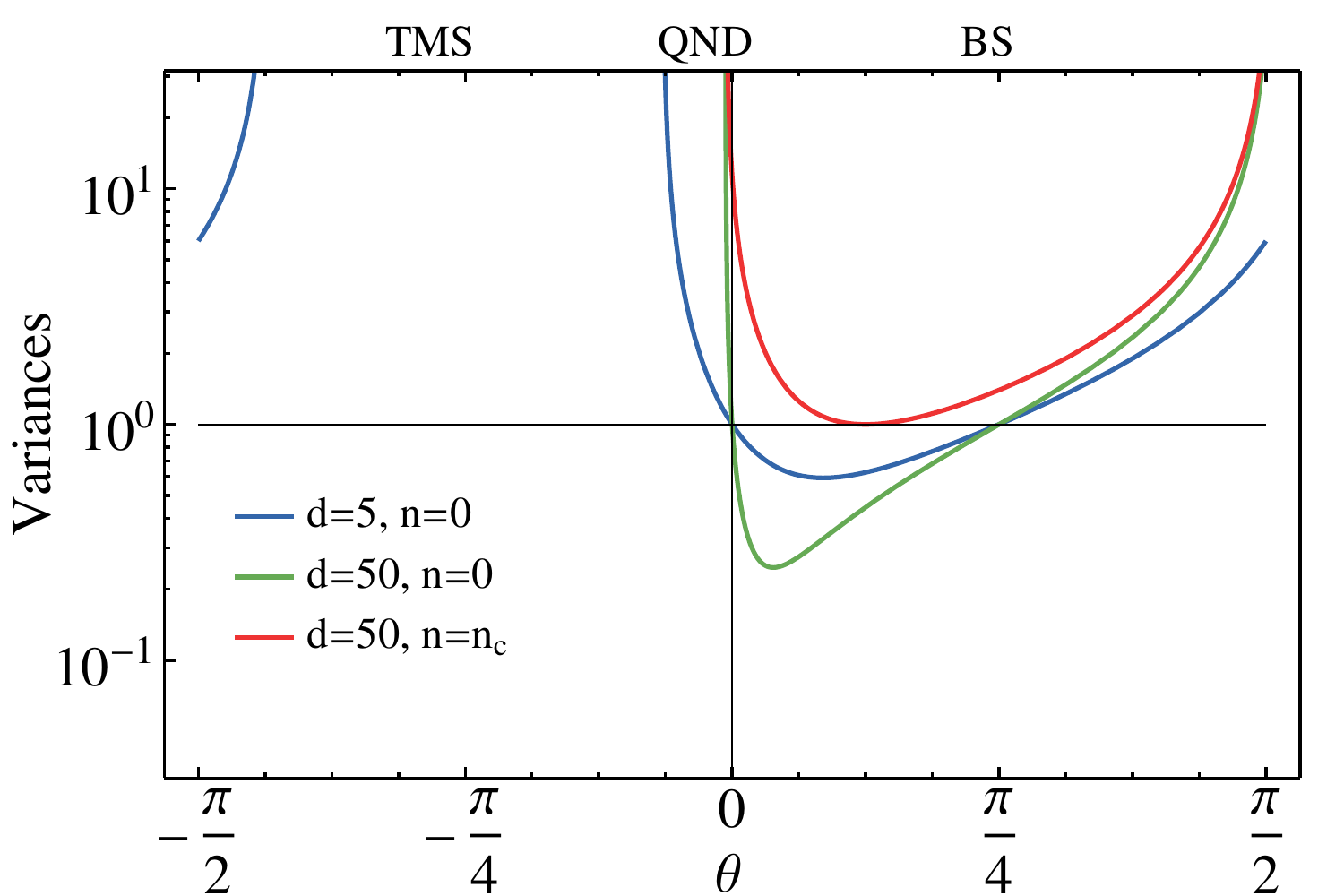}
   \caption{(Color online) Unconditional squeezed variance $V_\mathrm{u}$ (Eq.~\eqref{eq:VuODtheta}) versus interaction parameter $\theta$. The dynamics is stable only for $\theta>\theta_{c}$ where $\theta_c$ is given in \eqref{eq:thetac}. $V_\mathrm{u}$ exhibits a minimum for an optimal choice of $\theta^\mathrm{opt}_\mathrm{u}$ given in \eqref{eq:OptimalThetau} in a region where the beam splitter (BS) interaction dominates over the two mode squeezing (TMS) interaction. Curves for the following parameter values are shown: $d=5,\,n=0$ is the blue (dark grey) line, $d=50,\,n=0$ is the green (light grey) line, and $d=50,\,n=n_\mathrm{c}$ is the red (top) line.}
   \label{fig:Vu}
\end{figure}
\begin{figure}[t] 
   \centering
   \includegraphics[width=0.7\columnwidth]{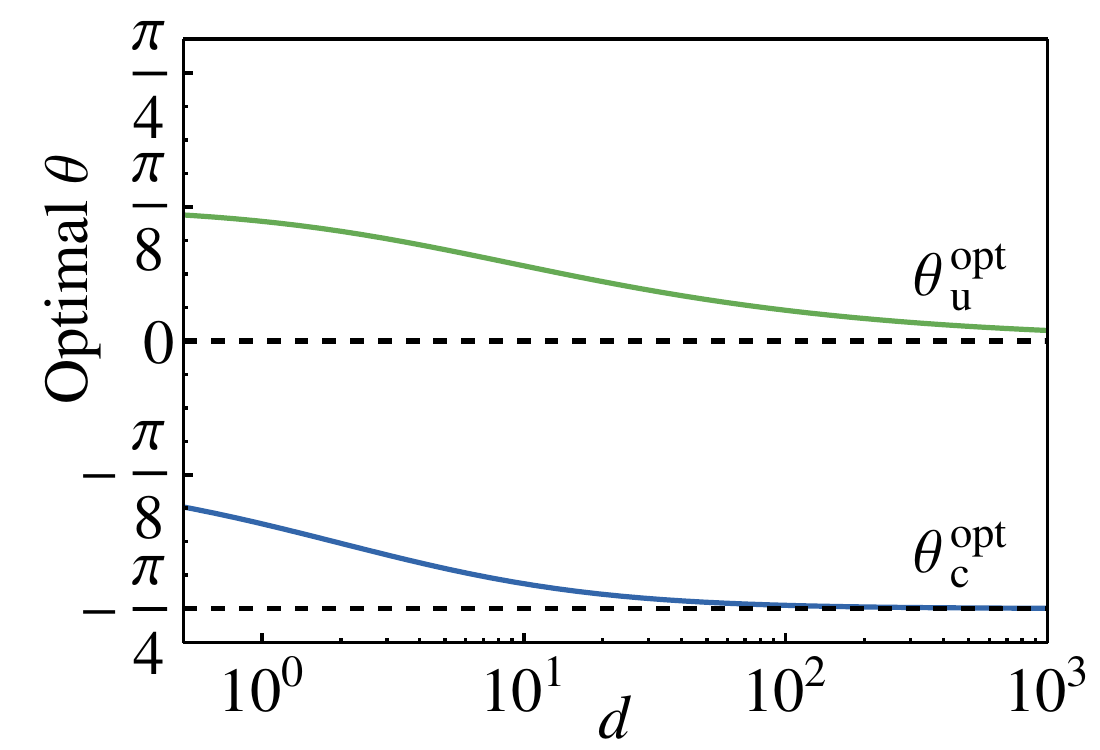}
   \caption{(Color online) Optimal interaction parameter $\theta$ versus optical depth. Solid curves  show conditional (the lower blue line) $\theta_\mathrm{c}^\mathrm{opt}$ and unconditional (the upper green line) $\theta_\mathrm{u}^\mathrm{opt}$ optimal interaction parameters. Dashed lines are asymptotics. For large optical depths the optimal interaction for dissipative generation of squeezing approaches the QND interaction ($\theta=0$), while the best choice for conditional generation of squeezing becomes the two mode squeezing interaction ($\theta=-\pi/4$).}
   \label{fig:optTheta}
\end{figure}
   \item
   Substituting the optimal interaction parameter in Eq.~\eqref{eq:VuODtheta} one obtains the minimal unconditional variance
   \begin{align}
   V^\mathrm{opt}_\mathrm{u} &= 2\frac{d\sqrt{(2n+1)(2n+1+d)+1}-d-2(2n+1)}{d^{2}-4}\notag\\
   \label{eq:uncondSQ}
   &\simeq 2\sqrt{\frac{2n+1}{d}}
   \end{align}
   where the approximation holds for $d\gg1$. Fig.~\ref{fig:varOD} below shows $V^\mathrm{opt}_\mathrm{u}$ versus optical depth. We recover the scaling of squeezing as $d^{-1/2}$ which also applies for a QND measurement. This can be seen directly from \eqref{eq:VcODtheta} by setting $\theta=0$ and $\epsilon=0$. In the asymptotic regime for large $d$ the QND squeezed variance is then $V_\mathrm{c}^\mathrm{QND}=\sqrt{(2n+1)/d}$, half of the dissipatively squeezed one.
   \end{itemize}

   \paragraph*{Conditional variance ($\epsilon=0$)}
   \begin{itemize}
   \item

   The conditional variance \eqref{eq:VcODtheta} exhibits a similar tradeoff. It has a minimum with respect to $\theta$, as shown in Fig.~\ref{fig:Vc}. The corresponding optimal interaction parameter $\theta^\mathrm{opt}_\mathrm{c}$ is defined in Appendix~\ref{appOptimalTheta} and shown in Fig.~\ref{fig:optTheta}, and in general corresponds to an interaction which is dominated by the two mode squeezing dynamics. It does not have a nice analytical form but in the limit of high optical depth can be expanded as
   \begin{equation}\label{eq:thetaCopt}
   \theta^\mathrm{opt}_\mathrm{c} \simeq -\frac{\pi}4 + \frac{2n+1}{d} + O(d^{-2}).
   \end{equation}
\begin{figure}[t] 
   \centering
   \includegraphics[width=0.9\columnwidth]{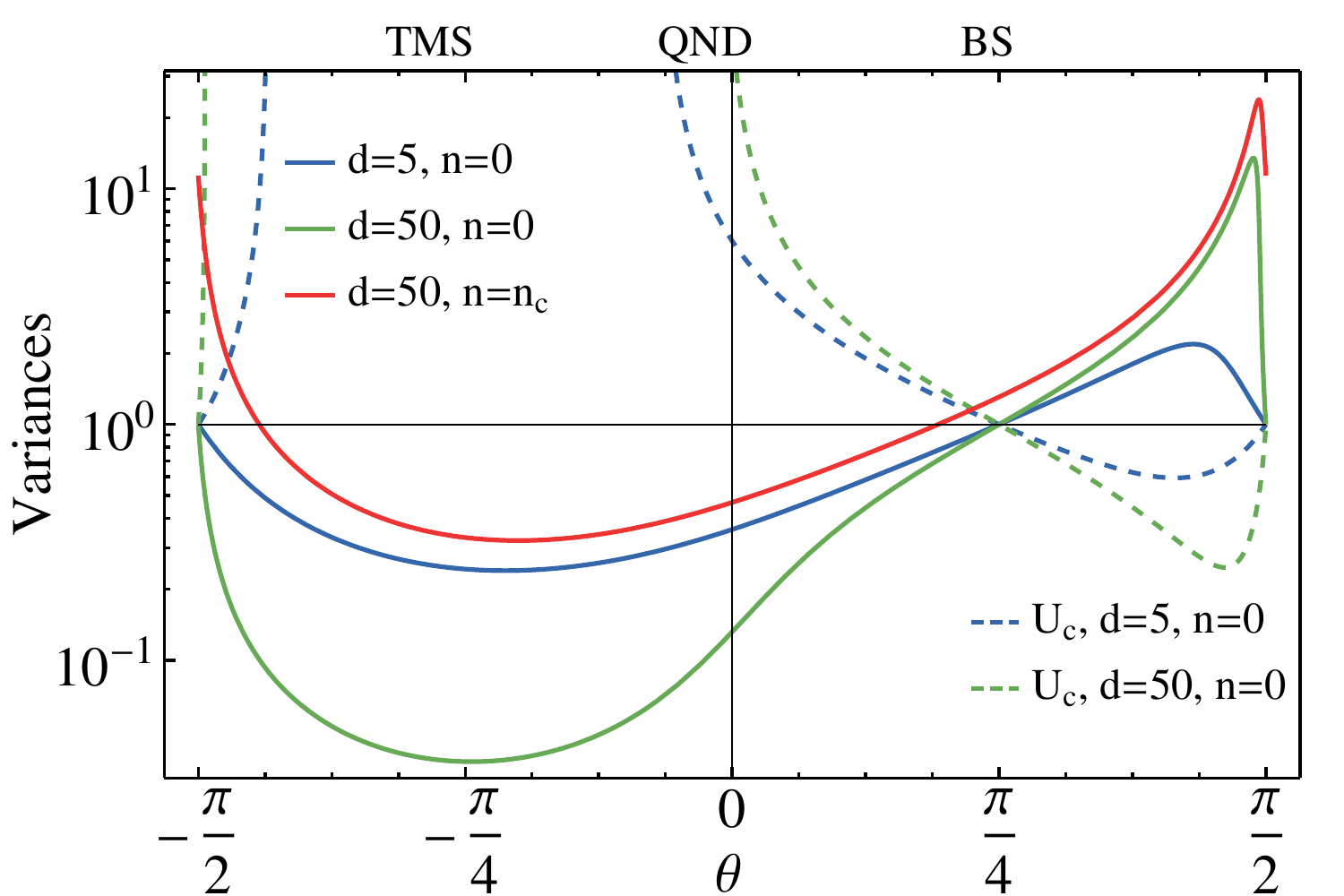}
   \caption{(Color online) Conditional squeezed variance $V_\mathrm{c}$ (solid) and unsqueezed variance $U_{c}$ (dashed) versus interaction parameter $\theta$. The beamsplitter reflectivity is $\epsilon=0$. Curves for the following parameter values are shown: $d=5,\,n=0$ is the blue (dark gray) solid line for $V_\mathrm{c}$ and dashed for $U_\mathrm{c}$, $d=50,\,n=0$ is the green (light grey) solid line for $V_\mathrm{c}$ and dashed for $U_\mathrm{c}$, and $d=50,\,n=n_\mathrm{c}$ is the red (top) solid line for $V_\mathrm{c}$.}
   \label{fig:Vc}
\end{figure}
   \item
   From the expression \eqref{eq:thetaCopt} one can see that the optimal interaction for the maximum entanglement is close to the two-mode squeezing regime. The expression for the optimised conditional variance in the limit of $d\gg1$ reads
   \begin{align}\label{eq:condSQ}
   V^\mathrm{opt}_\mathrm{c} &\simeq {\frac{2(2n+1)}{d}}.
   \end{align}
The striking feature of the conditional variance \eqref{eq:condSQ} is the inverse scaling with the optical depth in contrast to the $\sim1/\sqrt{d}$ scaling of the unconditional \eqref{eq:uncondSQ} squeezing (entanglement) for dissipative or QND generation for squeezing. The price we pay for this significantly improved scaling is the unbounded variance of the unsqueezed component. In the limit of $\epsilon\to 0$ the expression for the conditional variance $U_\mathrm{c}$ \eqref{eq:Uc} converges to the unconditional unsqueezed variance $U_\mathrm{u}$ \eqref{eq:Uu} and therefore too becomes unstable for $\theta\leq\theta_{c}$.
	\end{itemize}

\begin{figure}[t]
	\centering
  {\includegraphics[width=0.9\columnwidth]{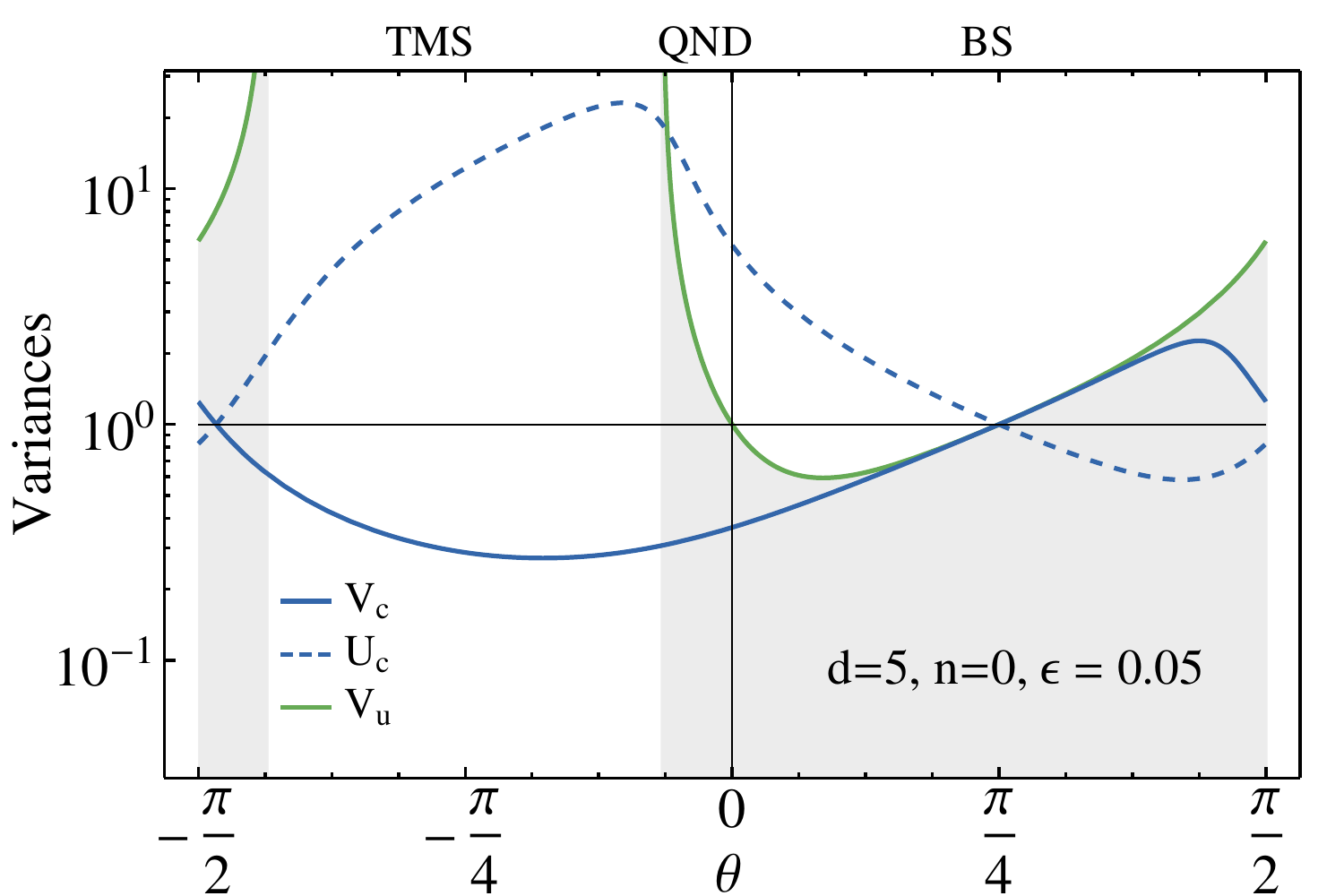}}
  \caption{(Color online) Squeezed unconditional variance $V_\mathrm{u}$ (green (light grey)), conditional variance $V_\mathrm{c}$ (blue (dark grey)) and unsqueezed conditional variance $U_\mathrm{c}$ (blue (dark grey) dashed) versus interaction parameter $\theta$ for an optical depth $d=5$, minimal atomic noise $n=0$ and beamsplitter reflectivity $\epsilon=0.05$. Interaction parameter $\theta=0$ corresponds to a QND interaction, for $0<\theta<\tfrac{\pi}2$ the beam splitter interaction dominates, for $-\tfrac{\pi}2<\theta<0$ the two mode squeezing interaction is dominant. The greyed region corresponds to the unconditionally stable dynamics.}
  \label{fig1}
\end{figure}

\begin{figure}[t]
	\centering
  {\includegraphics[width=0.9\columnwidth]{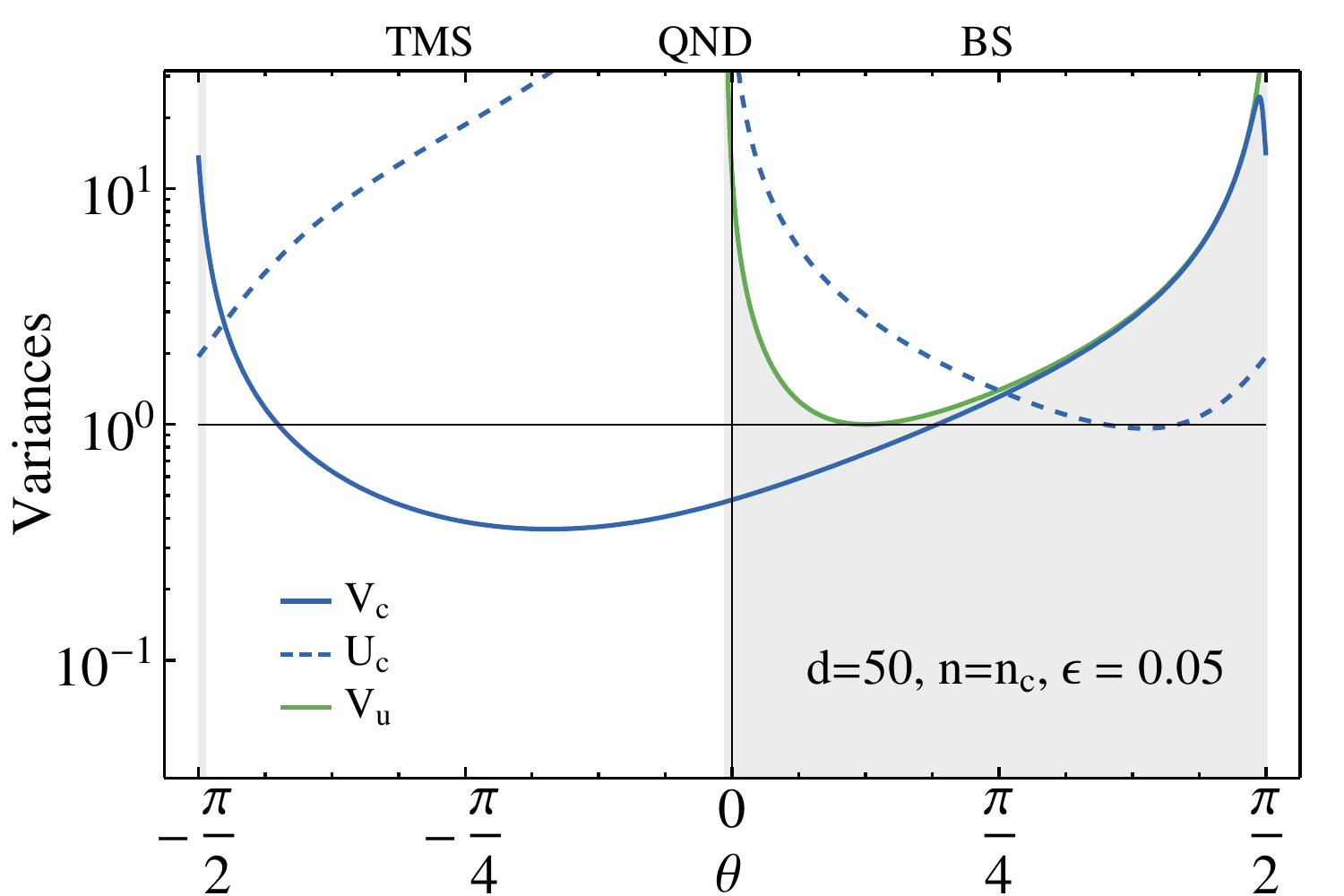}}
  \caption{(Color online) Same as for Fig.~\ref{fig1} but for critical (effective) temperature corresponding to decay to a thermal state with mean occupation number $n=n_{c}=d(\sqrt2-1)/4$. It is impossible to unconditionally achieve spin squeezing or entanglement for this or higher occupation number.}\label{fig2}
\end{figure}

	\paragraph*{Stabilised conditional variance ($\epsilon>0$)}
	\begin{itemize}
	\item
In order to profit from the enhanced scaling of squeezing with the optical depth the antisqueezed spin component therefore has to be actively stabilized via feedback. To do so it is necessary to detect also the second, conjugate light quadrature, from which the required information about the antisqueezed spin component can be gained. This can be achieved in the measurement setup shown in Fig.~\ref{fig:setup}. For a reflectivity $\epsilon>0$ a little bit of output light is substracted and subject to a homodyne detection of the amplitude quadrature. With appropriate feedback of the corresponding photocurrent the antisqueezed quadrature can be stabilized to a value which scales as $U_{c}\simeq 1/{\epsilon}$ (in units of shot or projection noise), at the cost of a somewhat reduced squeezing of the other spin component.  The variances for a small reflectivity $\epsilon$ are show in the Fig.~\ref{fig1} and Fig.~\ref{fig2}.

	\item
Most importantly, in Fig.~\ref{fig:varOD} we see that the $1/d$ scaling of squeezing can be stabilized for optical depths up to a value $d_{\ast}\simeq (2n+1)/\epsilon$. For greater optical depths the scaling of squeezing is changed back to the inverse square root law $V_\mathrm{c}^\mathrm{opt}(d>d_{\ast})\simeq 2\sqrt{(2n+1)\epsilon/d}$. The important conclusion is that for a given optical depth $d$ one can achieve an enhanced scaling of squeezing as $1/d$ by using a reflectivity $\epsilon \simeq (2n+1)/d$. The corresponding level of antisqueezing will be given by $U_\mathrm{c}(d_{\ast})\simeq\sqrt2/\epsilon\propto d$.

	\item
In the regime of unconditionally stable dynamics (shown by a grey area in Fig.~\ref{fig1} and Fig.~\ref{fig2}) the conditional state is dynamically stable too. Feedback may be applied in order to convert the unconditional squeezing into conditional squeezing. Depending on the application it might not be necessary to do so, and the conditionally squeezed state might give the same result (\emph{e.g.} in some Quantum Information protocol). On the contrary, the unstable regimes where the $1/d$ scaling of squeezing can be expected can be accessed only by applying a continuous feedback. The master equation describing the unconditional state of the system with feedback is given in Appendix \ref{appFeedback}.
   \end{itemize}

\begin{figure}[t]
	\centering
   {\includegraphics[width=0.9\columnwidth]{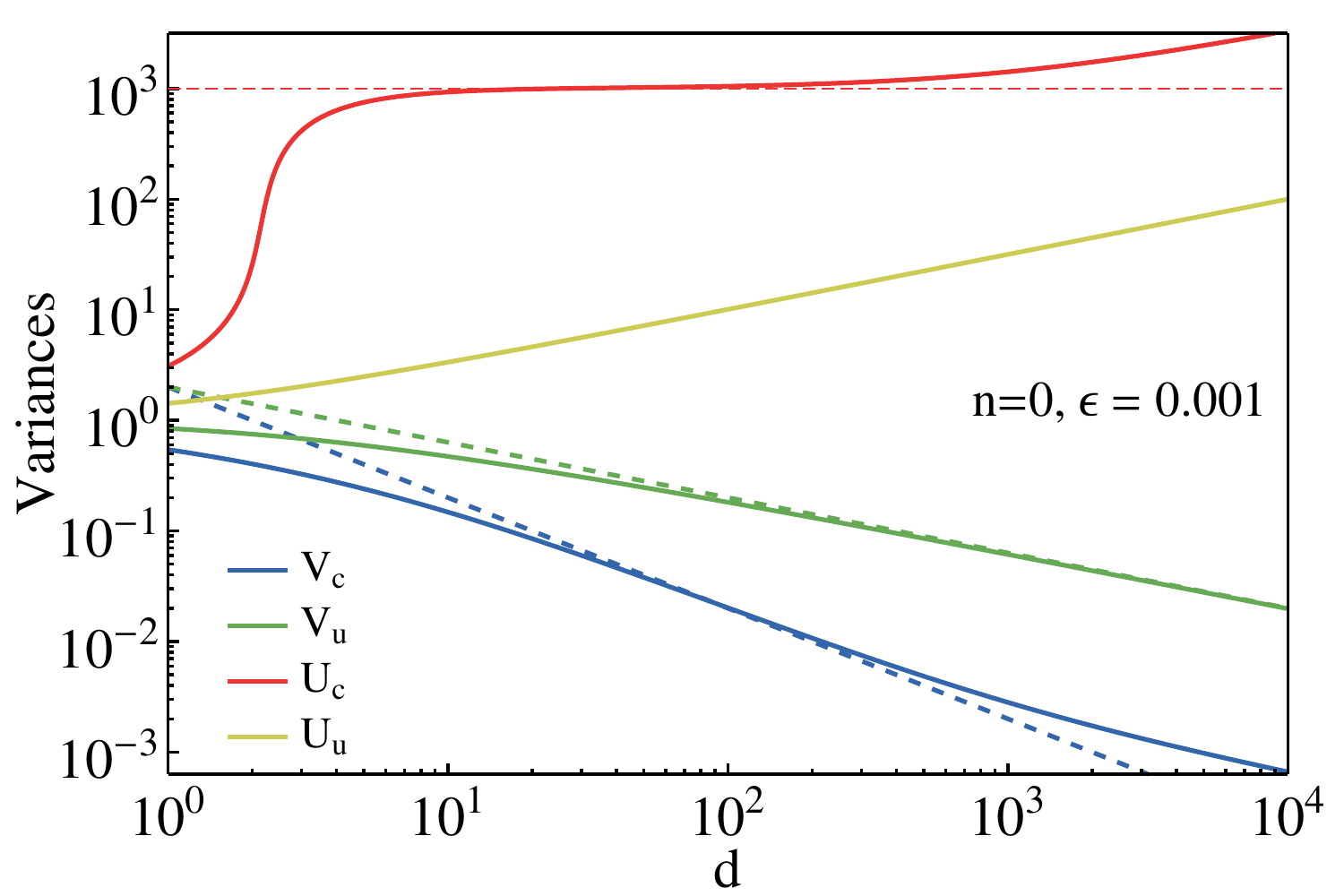}}
  \caption{(Color online) Squeezed and unsqueezed variances vs optical depth for minimal noise $n=0$: (blue (dark grey)) Conditional variance for the optimal $\theta^\mathrm{opt}_\mathrm{c}$, (green (lower light grey)) Unconditional squeezed variance for optimised interaction parameter $\theta^\mathrm{opt}_\mathrm{u}$, (yellow (upper light grey)) Unconditional unsqueezed variance for optimised interaction parameter $\theta^\mathrm{opt}_\mathrm{u}$, (red (upper)) Conditional unsqueezed variance for the optimal $\theta^\mathrm{opt}_\mathrm{c}$. Dashed lines: (blue (dark grey)) Asymptotic scaling $2/d$ for conditional dynamics and (green (light grey)) $2/\sqrt{d}$ for the unconditional dynamics, (red (upper)) asymptotic of the conditional unsqueezed variance $1/\epsilon$. }
  \label{fig:varOD}
\end{figure}


   \paragraph*{Finite occupation number}
   If the decay does not go back to the ground state (fully polarized state) we can identify an effective temperature giving rise to a mean thermal occupation $n$. Fig.~\ref{fig2} shows the effect on the unconditional and conditional variances. For a given optical depth $d$ there is a critical value of $n$ which should not be surpassed in order to achieve an entangled state.
   \begin{itemize}
   \item
   The unconditional steady state squeezing ($V_\mathrm{u}<1$) can only be observed if the occupation number of the environment is below a critical limit $n_{c} = d(\sqrt2-1)/4$.
	\item
    The conditional state can remain squeezed (entangled) until the temperature reach $\tilde n_{c}\sim \tfrac{5d}8 - \tfrac{3}{80d} + O(d^{-3})$ which is approximately six times higher than the unconditional $n_{c}$.
    \item
    Yet another critical temperature for the conditional squeezing in the dynamically stable regime is given by ${\tilde{\tilde n}}_{c}\sim\tfrac12+\tfrac{d}2-\tfrac{7}{16d}+O(d^{-3})$.
    \end{itemize}


 \section*{Conclusion}
 We have considered spin squeezing and entanglement of collective atomic spins obtained in a dissipative dynamics. We performed the optimisation over all possible quadratic light-matter interactions in a Gaussian description. The optimal unconditional squeezing exhibits a scaling with optical depth as $d^{-1/2}$. We found that if one performs homodyne measurements of the probe light it is possible to achieve better squeezing with a remarkable scaling as $d^{-1}$ by applying a feedback.

 Our results are based on a Gaussian description of the light matter interaction, and neglect the spatial extension of the atomic ensembles along the propagation direction. This is an excellent approximation for a QND interaction, but becomes less appropriate for dominant beam splitter or two mode squeezing interactions. However, we expect that our approximation still applies for ensembles at room temperature where thermal motion averages out spatial inhomogeneities. In view of the possible gain in squeezing a more careful treatment based on integration of Maxwell-Bloch equations is certainly desirable.

\begin{acknowledgments}
We acknowledge support from the EU project MALICIA, funding through the Centre for Quantum Engineering and Space-Time Research (QUEST) at the Leibniz University Hanover. The project MALICIA acknowledges the financial support of the Future and Emerging Technologies (FET) programme within the Seventh Framework Programme for Research of the European Commission, under FET-Open grant number: 265522. The figures for this article have been created using the LevelScheme scientific figure preparation system \cite{Caprio2005}.
\end{acknowledgments}

\begin{appendix}

\section{Stochastic Master Equations and Covariance Matrix Equations}\label{app:belavkin}

Consider a system composed of $a$ bosonic modes which obeys a stochastic Master equation
\begin{align}\label{eq:SMEQ}
\d\rho=-i[H,\rho]\d t+\sum_{i=1}^b \mathcal{D}[J_i]\rho\d t+\sum_{i=1}^c\mathcal{H}[J_i]\d W_i,
\end{align}
where $\mD[x]\rho=x\rho x^\dagger -\tfrac12\{x^\dagger x,\rho\}_+$ and $\mH[x]\rho=(x-\langle x\rangle)\rho+\rho(x^\dagger-\langle x^\dagger\rangle)$. For a $2a$ vector of canonical operators $\vec{r}^{\,T}=(x_1,p_1,\ldots,x_a,p_a)$ the Hamiltonian
\begin{align*}
  H&=\frac{1}{2}\vec{r}^{\,T}M\vec{r},
\end{align*}
is characterized by a real symmetric $2a\times 2a$ matrix $M$, and the $b$ jump operators
\begin{align*}
 J_i=\vec{r}^{\,T}\vec{j}_i,
\end{align*}
are determined by $b$ vectors $\vec{j}_i\in \mathds{C}^{2a}$. We assume that $c$ out of the $b$ decay channels are monitored, as described by the measurement terms proportional to the Wiener increments $\d W_i$. We restrict attention to the case where $\mathrm{d}W_{i}$ are independent Wiener processes $\d W_i\d W_j=\delta_{ij}\d t$.

If the state of the system is Gaussian it is fully described by the vector of first moments collected in the real $2a$-component \emph{displacement vector} $\vec{s}$, and the real symmetric $2a\times 2a$ covariance matrix $\Gamma$ defined by
\begin{align}\label{eq:Gaussian}
  s_i&=\mathrm{tr}\left\{\rho r_i\right\}, & \Gamma_{ij}&=\mathrm{tr}\left\{\rho (r_ir_j+r_jr_i)\right\}-2s_is_j,
\end{align}
such that for a single system and in the notation used in the main text
	\begin{equation}
	\Gamma = \begin{pmatrix}
				2V_\mathrm{c} & C_\mathrm{c}\\
				C_\mathrm{c}  & 2U_\mathrm{c}
	\end{pmatrix}.
	\end{equation}

It is then possible to show that for Gaussian states the equations of motion for $\vec{s}$ and $\Gamma$ implied by \eqref{eq:SMEQ} are
\begin{align}
  \d \vec{s} &= Q\vec{s}\,\d t + (\Gamma \vec{A} - \sigma \vec{B})\d \vec{W}, \notag\\
  \dot \Gamma &= \left(Q + 2\sigma \vec{B}\vec{A}^{\,T}\right)\Gamma + \Gamma \left(Q^T + 2\vec{A}\vec{B}^{\,T}\sigma^T\right)\notag\\
  &\quad+\left(P - 2\sigma\vec{B}\vec{B}^{\,T}\sigma^{T}\right) - 2\Gamma \vec{A}\vec{A}^{\,T} \Gamma.
  \label{eq:Belavkin}
\end{align}
The real matrices which enter here are determined from $M$ and $\vec{j}_i$ as follows:
\begin{align*}
  Q &= \sigma(M+R), & R&=-\frac{i}{2}\sum_{i=1}^b(\vec{j}_i^*\vec{j}_i^{\,T}-\mathrm{h.c.})\\
  P &= 2\sigma S\sigma^{T}, & S&=\frac{1}{2}\sum_{i=1}^b(\vec{j}_i^*\vec{j}_i^{\,T}+\mathrm{h.c.})\\
  \vec{A}&=\frac{1}{2}\sum_{i=1}^c(\vec{j}_i+\vec{j}_i^*), & \vec{B}&=-\frac{i}{2}\sum_{i=1}^c(\vec{j}_i-\vec{j}_i^*),
\end{align*}
and finally, $\sigma$ is the $2a\times 2a$ symplectic matrix
\[
\sigma=\bigoplus_{i=1}^a\begin{pmatrix} 0 & 1 \\-1 & 0 \end{pmatrix}.
\]
This can be proven by substituting \eqref{eq:SMEQ} into \eqref{eq:Gaussian}, using the cyclicity of the trace, the canonical commutator $[r_i,r_j]=i\sigma_{ij}$, and the property of Gaussian states that
\begin{multline*}
\mathrm{tr}\left\{\rho\{r_l,\{r_m,r_n\}_+\}_+\right\}=\\
2\left(\Gamma_{lm}s_n+\Gamma_{nl}s_m+\Gamma_{mn}s_l+2s_ns_ns_m\right).
\end{multline*}

\vspace{1cm}

\section{Optimal $\theta$}\label{appOptimalTheta}

An optimisation of the expression for the unconditional squeezing \eqref{eq:VuODtheta} with respect to the interaction parameter provides the optimal $\theta^\mathrm{opt}_\mathrm{u}$ given by
	\begin{align}
	\theta^\mathrm{opt}_\mathrm{u} &= \arctan\left[\frac{\sqrt{(2n+1)(2n+1+d)+1}-1}{2n+1+d}\right]\notag\\
	&\simeq \sqrt{\frac{2n+1}d}.
	\end{align}
The approximation assumes $d\gg1$.

The conditional variance \eqref{eq:VcODtheta} optimisation boils down to finding a proper root of a cubic equation:
	\begin{align*}
	0 &= 2x^{3}(1-2\epsilon)(2n+1+d\epsilon) \\
	&-x^{2}\left\{4(2n+1)^{2}-\epsilon[16n(n+1)+d^{2}]\right.\\
	&+ \left. d(2n+1)[2-(1-2\epsilon)^{2}]\right\}\\
	&- 2x(1-2\epsilon)(2n+1-d\epsilon)\\
	&+ d(2n+1)(1-2\epsilon).
	\end{align*}
The optimal interaction parameter is then given by $\theta^\mathrm{opt}_\mathrm{c} = \arctan(x)$.

\section{Feedback Master Equation}\label{appFeedback}

Based on the light measurement outcome one can apply a Hamiltonian feedback to the oscillator or to the EPR modes of a pair of them. The feedback Hamiltonian is proportional to the measured photocurrents $I_{1}(t)$ and $I_{2}(t)$ of the homodyne detectors,
	\begin{equation}
	H_\mathrm{fb} = I_{i}(t)F_{i},
	\end{equation}
where the feedback operators are $F_{1} = \xi_{1}P$ and $F_{2} = \xi_{2}X$ with $\xi_{1}$ and $\xi_{2}$ being the feedback gains.	
The resulting master equation reads
	\begin{align}\label{eq:fbME}
	\dot\rho = &-\frac{i}2[\{\sqrt{g(1-\epsilon)}F_{1} + i\sqrt{g\epsilon} F_{2}\}s +h.c.,\rho]\notag\\
	&+ \mD[\sqrt{g(1-\epsilon)}s-iF_{1}]\rho + g\epsilon\mD[\sqrt{g\epsilon}s-F_{2}]\rho,
	\end{align}
By choosing appropriate feedback gains $\xi_{1}$ and $\xi_{2}$ one can achieve steady state of the system with variances given by the conditional dynamics \eqref{eq:Vc} and \eqref{eq:Uc} for any interaction parameter~$\theta$. Applying Eqs.~\eqref{eq:Belavkin} to the master equation \eqref{eq:fbME} we arrive at the Lyapunov equations for the variances $V_\mathrm{fb}$ and $U_\mathrm{fb}$:

	\begin{align}
	\dot V_\mathrm{fb} = &- \left[\gamma + g\alpha\beta - 2\alpha \xi_{1} \sqrt{2g(1-\epsilon) }\right] V_{\mathrm{fb}} \notag\\
   &+\gamma(2n+1) + g\beta^2 - 2 \beta  \xi_{1}  \sqrt{2g(1-\epsilon) }+2 \xi_{1}^2,\\
   \dot U_\mathrm{fb} = &- \left[\gamma + g\alpha\beta - 2\beta \xi_{2} \sqrt{2g\epsilon}\right] U_{\mathrm{fb}} \notag\\
   &+\gamma(2n+1) + g\alpha^2 - 2\alpha\xi_{2}\sqrt{2g\epsilon}+2 \xi_{2}^2.
	\end{align}

\end{appendix}

\bibliography{GaussianEntanglement,ContinuousBellMeasurement}		

\end{document}